\documentclass[aps,twocolumn,epsfig,graphics,showpacs,floatfix,mathbbm]{revtex4}

\usepackage{amsmath,amsfonts,graphics,graphicx,epsfig,times,bbm}
\usepackage{amsfonts,graphics,times,bbm,amssymb}
\usepackage{epsfig}
\usepackage{graphicx}
\usepackage{color}
\usepackage{latexsym}
\usepackage{amscd}
\usepackage[all]{xy}

\begin{document}
\bibliographystyle{apsrev}

\def\setminus{\smallsetminus}
\def\qed{$\Box$}
\def\setminus{\smallsetminus}
\def\sem#1{[\hspace{-.35ex}[#1]\hspace{-.35ex}]}
\def\ens#1{\{#1\}}
\def\ie{\textit{i.e.}}
\def\eg{\textit{e.g.}}
\def\aka{\textit{a.k.a.}}
\def\iow{\emph{i.o.w.\ }}
\def\etc{\emph{etc\ }}
\def\st{^\star}
\def\dag{^\dagger}
\def\tr#1{\textcolor{red}{#1}}
\def\mathvec#1{\mathbf{#1}}
\def\al{\alpha}
\def\ba{\beta} 
\def\prd{\pi}
\def\rar{\rightarrow}
\def\lar{\leftarrow}
\def\srar{\mathrel{\rar\hskip-1.95ex\rar}}
\def\Rar{\Rightarrow}
\def\Lar{\Leftarrow}
\def\lrar{\longrightarrow}
\def\lRar{\Longrightarrow}
\def\Lrar{\Leftrightarrow}
\def\lRar{\leftrightarrow}
\def\ttt{\texttt} 
\def\mtt{\mathtt} 
\def\tsc{\textsc}
\def\tsf{\textsf} 
\def\msf{\mathsf} 
\def\mbf{\mathbf} 
\def\tbf{\textbf} 
\def\mcl{\mathcal} 
\def\mbb{\mathbb} 
\def\mfrk{\mathfrak} 
\def\mfr{\mathfrak}
\newtheorem{prop}{Proposition}\def\PRO{\begin{prop}}\def\ORP{\end{prop}}
\newtheorem{coro}{Corollary}\def\COR{\begin{coro}}\def\ROC{\end{coro}}
\newtheorem{theo}{Theorem}\def\TH{\begin{theo}}\def\HT{\end{theo}}
\newtheorem{defi}[prop]{Definition}\def\DE{\begin{defi}}\def\ED{\end{defi}}
\newtheorem{lemme}[prop]{Lemma}\def\LE{\begin{lemme}}\def\EL{\end{lemme}}
\newcommand{\AR}[2][c]{$$\begin{array}[#1]{lllllllllllllll}#2\end{array}$$}
\def\MA#1{\left(\begin{matrix}#1\end{matrix}\right)}
\def\EQ#1{\begin{eqnarray}#1\end{eqnarray}}
\def\hil#1{\mfr H_{#1}}
\def\hilo#1{\mfr H^1_{#1}} % unit vectors
\def\tup#1{\langle#1\rangle}
\def\ket#1{{|}#1\rangle}
\def\bra#1{\langle#1{|}}
\def\ctR{\mathop{\wedge}\hskip-.4ex} % controlled operators
\def\ctwo{{\mbb C}^2}  % qubits
\def\ctwoo{\mbb C^2_1} % norm 1
\def\ztwo{{\mbb Z}_2}
\def\ost{\frac1{\sqrt2}}
\def\most{\frac{-1}{\sqrt2}}
\def\pit{\frac\pi2}
\def\Cx#1{\cx{#1}{}}
\def\Cz#1{\cz{#1}{}}
\def\cz#1#2{Z_{#1}^{#2}}
\def\cx#1#2{X_{#1}^{#2}}
\def\ei#1{e^{i#1}}
\def\emi#1{e^{{-i}#1}}
\def\cx#1#2{X_{#1}^{#2}}
\def\phZ#1#2{N_{#1}^{#2}}
\def\phX#1#2{X_{#1}^{#2}}
\def\mLR#1#2#3#4{{}_{#4}[{M}_{#2}^{#1}]^{#3}}
\def\mR#1#2#3{\mLR{#1}{#2}{#3}{}}
\def\mL#1#2#3{\mLR{#1}{#2}{}{#3}}
\def\m#1#2{{M}_{#2}^{#1}}
\def\M#1#2{{M}_{#2}^{#1}}
\def\et#1#2{E_{#1#2}}
\def\CO#1{A_{#1}}
\def\ss#1#2{S_{#1}^{#2}}
\def\nf#1{\msf{nf}(#1)}
\def\brA#1{\rar_{#1}}
\def\mybox#1{\framebox{$#1$}}

\title{Determinism in the one-way model}

\author{Vincent Danos}
\affiliation{Universit\'e Paris~7 \& CNRS, 175 Rue du Chevaleret, 75013 Paris, France}
\author{Elham Kashefi}
\affiliation{Christ Church College, University of Oxford, OX1 1DP, Oxford, UK}

\date{\today}

\begin{abstract}
We introduce a \emph{flow condition} on open graph states (graph states with inputs and outputs) which guarantees globally deterministic behavior of a class of measurement patterns defined over them. Dependent Pauli corrections are derived for all such patterns, which equalize all computation branches, and only depend on the underlying entanglement graph and its choice of inputs and outputs. 

The class of patterns having flow is stable under composition and tensorization, and has unitary embeddings as realizations. The restricted class of patterns having both flow and reverse flow, supports an operation of adjunction, and has all and only unitaries as realizations. 
\end{abstract}

\pacs{03.67.Lx, 03.67.-a, 03.67.Mn}

\maketitle

\section{Introduction}
The recent one-way quantum computing model~\cite{RB01,RB02,mqqcs} has already drawn considerable attention, because it suggests different physical realizations of quantum computing~\cite{Nielsen04,CMJ04,BR04,TPKV04,TPKV05,nature05,KPA05,BES05,CCWD05}. However, whether this fundamentally different model may also suggest new insights in quantum information processing still stands as an open question.

Computation in this model, consists of a first phase of preparation and entanglement, followed by 1-qubit measurements and a final round of corrections. Making measurements an integral part of computation will in general induce non-deterministic behaviors. To counter this, both measurements and corrections are allowed to depend on the outcomes of previous measurements. This mechanism of feed-forwarding classical observations is known to be a necessary requirement for the model to be universal~\cite{DKP04}. Whether and how a given pattern can be controlled so as to obtain a globally deterministic behavior is the question we address in this paper. 

A variety of methods for constructing measurement patterns have been already proposed \cite{mqqcs, graphstates, CLN04}  that guarantee determinism by construction. We introduce a direct condition on open graph states (graph states with inputs and outputs) which guarantees a strong form of deterministic behavior for a class of one-way measurement patterns defined over them.  Remarkably, our condition bears only on the geometric structure of the entangled graph states. This condition singles out a class of patterns with flow, which is stable under sequential and parallel compositions and is large enough to realize all unitary and unitary embedding maps.

Patterns with flow have interesting additional properties. First, they are uniformly deterministic, in the sense that no matter what the measurements angles are, the obtained set of corrections, which depends only on the underlying geometry, will make the global behavior deterministic. Second, all computation branches have equal probabilities, which means in particular these probabilities are independent of the inputs, and as a consequence, one can show that all such patterns implement unitary embeddings. Third, a more restricted class of patterns having both flow and reverse flow supports an operation of adjunction, corresponding to time-reversal of unitary operations. This smaller class implements all and only unitary transformations. Moreover, for open graph states with flow, one can derive a direct procedure for realization of unitaries as measurements patterns \cite{BDK06}.

\section{Measurement Patterns}

We briefly recall the definition of measurement patterns and various notions of determinism. More detailed introductions can be found in \cite{Nielsen05,Jozsa05,BB06}. In this paper, we will employ an algebraic approach called, the \emph{Measurement Calculus} \cite{DKP04}. Computations in a pattern involve a combination of 1-qubit preparations $\phZ i \al$, 2-qubit entanglement operators $\et ij:=\ctR Z_{ij}$ (controlled-$Z$), 1-qubit measurements $\M\al i$, and 1-qubit Pauli corrections $\Cx i$, $\Cz i$, where $i$, $j$ represent the qubits on which each of these operations apply, and $\al$ is a parameter in $[0,2\pi)$.  

Preparation $\phZ i \al$ prepares qubit $i$ in state $\ket{+_\al}_i$, where $\ket{\pm_{\al}}$ stand for $\ost(\ket0\pm\ei{\al}\ket1)$. Measurement $\M\al i$ is defined by orthogonal projections $\ket{\pm_\al}\bra{\pm_\al}_i$, applied at qubit $i$, with the convention that $\ket{+_\al}\bra{+_{\al}}_i$ corresponds to the outcome $0$, while $\ket{-_\al}\bra{-_{\al}}_i$ corresponds to $1$. Note that we consider here only destructive measurements, \ie\,a projection $\ket{\psi}\bra{\psi}$ is always followed by a trace out operator and hence we might write it as $\bra{\psi}$. 

Qubits are measured at most once, therefore we may represent unambiguously the outcome of the measurement done at qubit $j$ by $s_j$.  
Dependent corrections, used to control non-determinism, will be written $\cx i{s_j}$ and $\cz i{s_j}$, with $\cx i0=\cz i0=I$, $\cx i1=\Cx i$, and $\cz i1=\Cz i$. 

A \emph{measurement pattern}, or simply a pattern, is defined by the choice of $V$ a finite set of qubits, two possibly overlapping subsets $I$ and $O$ determining the pattern inputs and outputs, and a finite sequence of commands acting on $V$. 

Such a pattern is said to be \emph{runnable} if it satisfies the following: (R0) no command depends on an outcome not yet measured, (R1) no command acts on a qubit already measured or not yet prepared (except preparation commands), and (R2) a qubit $i$ is measured (prepared) if and only if $i$ is not an output (input).

Write $\hil I$ ($\hil O$) for the Hilbert space spanned by the inputs (outputs).
The run of a runnable pattern consists simply in executing each command in sequence. If $n$ is the number of measurements (which by (R2) is also the number of non outputs) then the run may follow $2^n$ different branches. Each branch is associated with a unique binary string  $\mathvec s$ of length $n$, representing the classical outcomes of the measurements along that branch, and a unique 
\emph{branch map} $A_{\mathvec s}$ representing the linear transformation from $\hil I$ to $\hil O$ along that branch. 

Branch maps decompose as $A_{\mathvec s}=C_{\mathvec s}\Pi_{\mathvec s}U$, where $C_{\mathvec s}$ is a unitary map over $\hil O$ collecting all corrections on outputs, $\Pi_{\mathvec s}$ is a projection from $\hil V$ to $\hil O$ representing the particular measurements performed along the branch, and $U$ is a unitary embedding from $\hil I$ to $\hil V$ collecting the branch preparations, and entanglements. Therefore
\AR{ 
\sum_{\mathvec s} A_{\mathvec s}\dag A_{\mathvec s}=\sum_{\mathvec s} U\dag\Pi_{\mathvec s}U=I
}
and $T(\rho):= \sum_{\mathvec s} A_{\mathvec s}\rho A_{\mathvec s}\dag$ is a trace-preserving completely-positive map (cptp-map), explicitly given as a Kraus decomposition. One says that the pattern \emph{realizes} $T$. 
 
A pattern is said to be \emph{{deterministic}} if it realizes a cptp-map that sends pure states to pure states. This is equivalent to saying that branch maps are proportional, that is to say, for all $q\in\hil I$ and all $\mathvec s_1$, $\mathvec s_2\in \ztwo^n$, $A_{\mathvec s_1}(q)$ and $A_{\mathvec s_2}(q)$ differ only up to a scalar. A pattern is said to be \emph{strongly deterministic} when branch maps are equal, \ie, for all $\mathvec s_1$, $\mathvec s_2\in \ztwo^n$, $A_{\mathvec s_1}=A_{\mathvec s_2}$. A pattern is said to be \emph{uniformly deterministic} if it is deterministic for all values of its measurement angles.

\LE
Strongly deterministic patterns realize unitary embedding maps.
\EL
{\em Proof.} 
If a pattern is strongly deterministic and realizes the map $T$ then
\AR{
T(\rho)=A\rho A\dag
} with $A:=2^{n/2}A_{\mathvec s}$, and $A$ must be a unitary embedding, because $\sum_{\mathvec s} A_{\mathvec s}\dag A_{\mathvec s}=A\dag A=I$. In such cases, one says that the pattern realizes the unitary embedding $A$. \qed

{\em Example.} Not all deterministic patterns are uniformly or strongly so. To see this, 
choose as command sequence $\cx 1{s_2}\m 02 \et 12\phZ 20$, with $V=\ens{1,2}$, and $I=O=\ens1$. The two branch maps are given by 
$A_0=\ket0\bra0$, and $A_1=\ket0\bra1$, so they are proportional, but distinct,
and the pattern is deterministic, but not strongly so. The associated cptp-map $T(\ket{\psi}\bra{\psi})=\tup{\psi,\psi}\ket0\bra0$ projects any state onto $\ket{0}$ and does not correspond to a unitary transformation. This pattern is not uniformly deterministic either, since $\al=0$ is the only angle value for $\m\al2$ which makes it deterministic.

\section{Geometries and Flows}

An \emph{open graph states} $(G,I,O)$ consists of an undirected graph $G$ together with 
two subsets of nodes $I$ and $O$, called inputs and outputs. We write $V$ for the set of nodes in $G$, $I^c$, and $O^c$ for the complements of $I$ and $O$ in $V$, $G(i)$ for the set of neighbors of $i$ in $G$, and $E_G:=
\prod_{(i,j)\in G}E_{ij}$ for the global entanglement operator associated to $G$.

One may think of an open graph state as the beginning of the definition of a pattern, where one has already decided how many qubits will be used ($V$), how they will be entangled ($E_G$), and which will be inputs and which outputs ($I$ and $O$). To complete the definition of the pattern it remains to decide which angles will be used to prepare qubits in $I^c$ (qubits in $I$ are given in an arbitrary states) which angles will be used to measure qubits in $O^c$, and most importantly, if one is interested in determinism, which dependent corrections will be used. Conversely, any pattern has a unique underlying open graph state, obtained by forgetting preparations, measurements and corrections. 

For instance, the open graph state associated to the example  above is the graph $G$ with nodes $\ens{1,2}$, inputs and outputs $\ens1$, and $E_G=\et 12$. To complete the definition, one has to choose the angles of the measurement and preparations done at qubit $2$, and define the dependent corrections.

We give a condition bearing on the geometry of open graph states, under which one can construct a set of dependent corrections such that the obtained pattern is strongly and uniformly deterministic. 

\DE An open graph state $(G,I,O)$ has \emph{flow} if there exists a map $f:O^c\rar I^c$ (from measured qubits to prepared qubits) and a partial order $>$ over $V$ such that for all $i\in O^c$:
\\--- (F0) $(i,f(i))\in G$
\\--- (F1) $f(i)>i$
\\--- (F2) for all neighbours of $f(i)$ except 
$i$ ($k\in G(f(i))\smallsetminus\ens i$), we also have $k>i$
\ED

As one can see, a flow consists of two structures: a function $f$ over vertices and a matching partial order over vertices. In order to obtain a deterministic pattern for an open graph state with flow, dependent corrections will be defined based on function $f$. The order of the execution of the commands is given by the partial order induced by the flow. The matching properties between the function $f$ and the partial order $>$ will make the obtained pattern runnable. 

\begin{figure}[h]
\begin{center}
\includegraphics[scale=0.5]{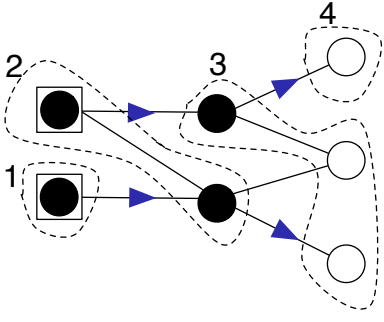}
\caption{An open graph state with flow. The boxed qubits are the inputs and white circles are the outputs. All the non-output qubits, black circles, will be measured during the run of the pattern. The flow function is represented as arrows and the partial order on the vertices are given by the 4 partition sets. }
\label{flowflow}
\end{center}
\end{figure}

Figure~\ref{flowflow} shows an open graph state together with a flow, where function $f$ represented as arrows from $O^c$ (measured qubits, black circles) to $I^c$ (prepared qubits, non boxed nodes). The associated partial order is given by the labeled sets of vertices. The coarsest order $>$ such that (F1) and (F2) holds is called the
\emph{dependency order} induced by the flow, and the number of the partition sets (4 in Figure~\ref{flowflow}) is called the \emph{depth} of the flow. In general flows may or may not exist, and are not unique either.

\TH \label{t-flow} Suppose the open graph state $(G,I,O)$ has flow $(f,>)$, then the pattern:
\AR{
\mfr P_{f,G,>,\vec\al}&:=&\prod^>_{i\in O^c}
(\cx{f(i)}{s_i}\prod_{k\in G(f(i))\setminus\ens{i}}\cz{k}{s_i}\m{\al_i}i)
E_G \phZ {I^c}0
}
where the product follows the dependency order $>$, is runnable, uniformly and strongly deterministic, and realizes the unitary embedding:
\AR
{U_{G,I,O,\vec\al}&:=&
(\prod_{i\in O^c}\bra{{+_{\al_i}}}_i)\, E_G\phZ {I^c}0}
\HT
{\em Proof.} The proof is based on the following equations, where $s$ stands for any arbitrary $s_j$:
\EQ{
\label{e1}
\bra{{+_\al}}_i&=&\m\al i\cz i{s_i}\\ 
\label{e2}
\cz i{s}\et ij&=& \cx j{s}\et ij\cx j{s}\\
\label{e3}
\cx i{s}\et ij&=&\et ij\cz j{s}\cx i{s}\\
\label{e4}
\cz i{s}\et ij&=&\et ij\cz i{s}\\
\label{e5}
\cx i{s}\phZ i0&=&\phZ i0
}
Equation~(\ref{e1}) amounts to saying that $Z_i\ket{\pm_\al}_i=\ket{\mp_\al}_i$; notice also that this property uniquely defines $Z$. Equations~(\ref{e2}), (\ref{e3}), and~(\ref{e4}) come from the fact that $\ctR Z$ is in the normalizer of the Pauli group, and are easy to verify. Equation (\ref{e5}) is obvious. From (\ref{e1}) we obtain:
\AR{
\prod_{i\in O^c}\bra{{+_\al}}_i E_G\phZ {I^c}0&=_{(\ref{e1})}&
(\prod_{i\in O^c}\m{\al_i}i\cz i{s_i})E_G\phZ {I^c}0
}
so the right hand side is clearly a deterministic pattern, but just as clearly it violates condition (R0), since $\cz i{s_i}$ depends on a measurement which has not been done yet. At that point, entanglement comes to rescue. Write $G(i)^c$ for the graph obtained by removing $G(i)$ from $G$. Then we can rewrite the above pattern as follows, where boxes represent the part to which we apply the rewriting equations:
\AR{
\cz i{s_i}E_{G}\phZ {I^c}0
&=&\\
\cz i{s_i}E_{G(i)}E_{G(i)^c}\phZ {I^c}0
&=&\\
\mybox{\cz i{s_i}\et i{f(i)}}
(\prod_{k\in G(f(i))\setminus\ens i})
\et{f(i)}kE_{G(f(i))^c}\phZ {I^c}0
&=_{(\ref{e2})}&\\
\cx {f(i)}{s_i}\et{i}{f(i)}
\mybox{\cx {f(i)}{s_i}
(\prod_{k\in G(f(i))\setminus\ens i}\et{f(i)}k})
E_{G(f(i))^c}\phZ {I^c}0
&=_{(\ref{e3})}&\\
\cx {f(i)}{s_i}
\mybox{
E_{G(f(i))}
(\prod_{k\in G(f(i))\setminus\ens i}\cz k{s_i})
} 
\cx {f(i)}{s_i}
E_{G(f(i))^c}\phZ {I^c}0
&=_{(\ref{e4})}&\\
\cx {f(i)}{s_i}
(\prod_{k\in G(f(i))\setminus\ens i}\cz k{s_i})
E_{G}
\mybox{
\cx {f(i)}{s_i}
\phZ {I^c}0}
&=_{(\ref{e5})}&\\
\cx {f(i)}{s_i}
(\prod_{k\in G(f(i))\setminus\ens i}\cz k{s_i})
E_{G}\phZ {I^c}0
}
Condition (F0) is used in the third step. Finally: 
\AR{
\prod_{i\in O^c}\bra{{+_\al}}_i E_G\phZ {I^c}0=\\(\prod_{i\in O^c}\cx {f(i)}{s_i}(\prod_{k\in G(f(i))\setminus\ens i}\cz k{s_i})
\m{\al_i}i)E_G\phZ {I^c}0
}
By conditions (F1) and (F2) the obtained pattern is runnable, since the product
can always be ordered according to $>$. Moreover, by the last equation, all branch maps are equal, and therefore the pattern is strongly deterministic. Finally, since the proof uses nowhere the
particular values of the measurement angles $\al_i$, it is also uniformly so.
\qed

The intuition of the proof is that Equation \ref{e2} converts an anachronical $Z$ correction at $i$, given in the term $M_i^\al Z_i^{s_i}$, into a pair of a `future' $X$ correction, the one sent to $f(i)$ (so in the future, by condition (F1)) and a `past' $X$ correction, sent to the past, until it reaches a preparation, where it is absorbed because of Equation \ref{e5}.

Note that the unitary embedding associated to $\mfr P_{f,G}$ (we drop $\vec\al$ and $>$, for simplicity) does not depend on the flow. Yet, the choice of $(f,>)$ determines the structure of the corrections used by the pattern and the order of the execution, and has therefore an influence on its depth complexity, which is the depth of the flow. 

Another thing worth noticing, is that using the graph stabilizer~\cite{NC00,graphstates} at $i$, defined as $K_{G(i)}:=X_{i}(\prod_{j\in G(i)}Z_j)$, the pattern $\mfr P_{f,G}$ can be equivalently written as:
\AR{
\mfr P_{f,G}&=&
\prod^>_{i\in O^c}
(
\m{\al_i}i \cz i{s_i}
K_{G(f(i))}^{s_i}
)
E_G \phZ {I^c}0
}
and the above proof can be reread in terms of stabilizers. In another word, for cancelling an anachronical $Z$ correction at $i$ it is enough to apply the dependent stablizer at qubit $f(i)$, $K^{s_i}_{G(f(i))}$ and again conditions $(F1)$ and $(F2)$ guarantee that the obtained pattern is runnable. 

\subsection{Pauli Measurements}

\begin{figure}[h]
\begin{center}
\includegraphics[scale=0.5]{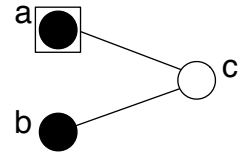}
\caption{An open graph state with no flow, since for any candidate function $f$ we have $f(a)=f(b)=c$ and therefore there exists no matching partial order as $a$ should be in future of $b$ and $b$ in future of  $a$. }
\label{noflow}
\end{center}
\end{figure}

As we saw before, not all open graph states have flow. Figure~\ref{noflow} shows such an example, let $f$ be a candidate flow function, then the only choice for $f(a)$ is node $c$, same is true for $f(b)$. Now from condition (F2) node $b$ must be in the future of node $a$ and vice versa. Hence we reach a causality conflict.  

\begin{figure}[h]
\begin{center}
\includegraphics[scale=0.5]{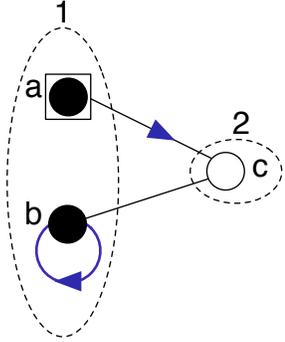}
\caption{An open graph state where the node with a loop, $b$ must be measured with a Pauli $Y$ measurement. The matching partial order has two levels given by the doted partitions.}
\label{loopflow}
\end{center}
\end{figure}

However, one can still obtain a deterministic pattern for the open graph state in Figure \ref{noflow} by fixing the angle of the measurement of node $b$ to be $\pi/2$. To see why, recall that Condition (F1) forbids $f(i)=i$, yet, in the special case where qubit $i$ is measured with angle $\pit$ (Pauli $Y$ measurement), choosing $f(i)$ to be $i$ will work, since:
\AR{
\m {\frac{\pi}{2}}i \cx is = \m{\frac{\pi}{2}}i \cz is 
}
Hence to correct the $Y$ measurement at qubit $i$ one can apply the dependent stabilizer, $(Z_i(\prod_{j\in G(i)}Z_j))^{s_i}$, at the same qubit $i$ instead of a neighboring qubit, Figure \ref{loopflow}. However the obtained pattern is  deterministic only if qubit $b$ is measured with angle $\pit$, and is therefore not uniformly deterministic. 

\begin{figure}[h]
\begin{center}
\includegraphics[scale=0.5]{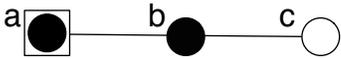}
\caption{An open graph sate where node $b$ will be measured with $Y$ measurement however one cannot apply the special case described above.}
\label{Ygraph}
\end{center}
\end{figure}

Note that in the above example we fixed $f(b)=b$ but condition $F(3)$ still need to be verified. And this is indeed the case since in the given partial order the qubit $c$ which is neighbour of qubit  $b$ is in the next level. To make this point clear consider the open graph state in Figure \ref{Ygraph}. 

\begin{figure}[h]
\begin{center}
\includegraphics[scale=0.5]{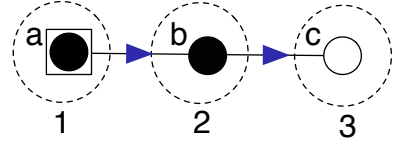}
\caption{An open graph sate where node $b$ will be measured with $Y$ with a corresponding flow without any loop.}
\label{Ygraphflow}
\end{center}
\end{figure}

The only choice for $f(a)$ is $b$ and hence $a<b$ but then letting $f(b)=b$ will violate the $F(3)$ condition. Therefore the only solution is to consider $Y$ measurement as an arbitrary measurement then we obtain a flow, Figure \ref{Ygraphflow}.  

Another special case is when qubit $f(i)$ is measured with angle $0$ (Pauli $X$ measurement). Again the requirement that $f(i)>i$ can be dropped because:
\AR{
\m 0i \cx is = \m 0i  
}
Therefore in the flow construction where the neighboring qubit $f(i)$ receives $X_{f(i)}^{s_i}$, if it is measured with angle $0$ this correction can be ignored. 

The special cases of Pauli measurements can be related to the fact that Pauli measurements transform one graph state to another one \cite{graphstates}. Hence one can observe that for open graph states without flow, there might exists a set of Pauli measurements that transform it to one with flow. 

\subsection{Circuit Decomposition}

\begin{figure}[h]
\begin{center}
\includegraphics[scale=0.5]{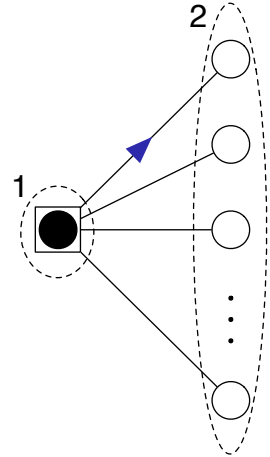}
\caption{Star pattern $\mfr S(n,\al)$ with one input, the boxed circle, and $n$ outputs, white circles. The input qubit will be measured with angle $\al$ and one of the outputs receives a dependent correction $X^{s_1}$. The flow is given by the single arrow from the input to one of the outputs and two level partial order.}
\label{extJ}
\end{center}
\end{figure}

\begin{figure}[h]
\begin{center}
\includegraphics[scale=0.4]{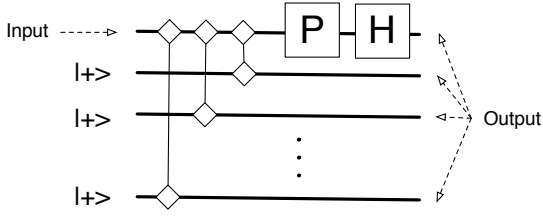}
\caption{The circuit implementation of Star pattern in Figure \ref{extJ} , with controlled-$Z$, phase $P(-\al)$ and Hadamard $H$ gates.}
\label{cirJ}
\end{center}
\end{figure}

Flow also provides a decomposition into simple building blocks, called \emph{star patterns}, from which one can derive a corresponding circuit implementation of the pattern. Define the star pattern $\mfr S(n,\al)$ as:
\AR{
&\cx 2{s_1}\M{{\al}}1\et 12\et 13 \cdots \et 1n
}
where 1 is the only input and $2,\cdots, n$ are the outputs, for $n\geq2$. The underlying graph has a simple flow function with $f(1)=2$ and a two level partial order (see Figure \ref{extJ}). It is easy to verify that the Star pattern implements the unitary given by the circuit  in Figure \ref{cirJ}.

Every pattern such that the underlying open graph state has flow can be decomposed into star patterns.
The construction starts by picking a qubit in the first level of the partial order, exhausts all qubits in the first level before going to the next level. Each time a qubit $i$ is picked the associated star pattern $S_i$ is taken to have $i$ as input, and all its remaining current neighbours as outputs. Then we remove this qubit from the graph and carry on the construction till we reach to the final level of partial order. The final deterministic pattern is the sequential and tensor composition of the obtained star patterns with the final $\ctR Z$ between the output qubits:
\AR{
\mfr P=\prod_{m,n \in O} \et mn\prod^>_{i\in O^c} S_i
}
Now each Start pattern can be replaced by its corresponding circuit to give a circuit decomposition for the pattern $\mfr P$. In the obtained circuit each wire represents either an input qubit or an auxillary one prepared in $\ket +$ state, where the case is determined during the above construction. This construction can be easily formalized. 
 
\section{Algebraic Structure}

As yet, Theorem \ref{t-flow} is only valid when preparations are all of the form $\phZ i0$ since Equation (\ref{e5}) in the proof is valid only for such preparations. 
Define $\phX i\al=\cz i\al \Cx i\cz i{-\al}$, with $Z_i^\al$ the phase operator 
with angle $\al$ applied at $i$. One has $Z_i=\cz i{\pi}$. 
To handle general phase preparations, one only needs the analog of equations (\ref{e2}), (\ref{e3}) and (\ref{e5}):
\AR{
\label{xe2}
\cz i{s}\et ij&=&(\phX i\al){^s}\et ij(\phX i\al){^s}\\
\label{xe3}
(\phX i\al)^{s}\et ij&=&\et ij\cz j{s}(\phX i\al){^s}\\
\label{xe5}
(\phX j\al)^{s}\phZ i\al&=&\phZ i\al
}
and now Theorem \ref{t-flow} works as before. Note that we had to extend the set of corrections to include $\phX i\al$. This extension will prove natural below, when we deal with adjunction. 

Say an open graph state $(G,I,O)$ has \emph{bi-flow}, if both $(G,I,O)$ and its dual state $(G,O,I)$ have flow. Say a pattern has flow (bi-flow) if its underlying open graph state does. 

The class of patterns with flows (bi-flows) is closed under composition and tensorization. It is also universal, in the sense that all unitaries can be realised within this class. This follows from the existence of a set of generating patterns having bi-flow~\cite{generator04}.

\begin{figure}[h]
\begin{center}
\includegraphics[scale=0.4]{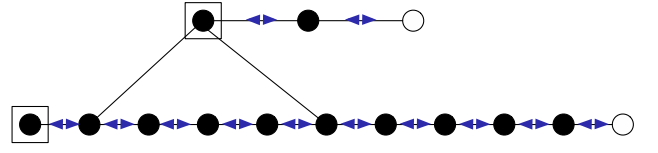}
\caption{An open graph state with bi-flow.}
\label{biflow}
\end{center}
\end{figure}

Figure~\ref{biflow} shows the open graph state corresponding to a pattern realizing
$\ctR U$ (controlled-$U$), for $U$ an arbitrary 1-qubit unitary ~\cite{generator04}.

Patterns with bi-flows realize unitary operators. Indeed, by (F2), a flow $(f,>)$
is one-to-one. Therefore the orbits $f^n(i)$ for $i\in I$ define an injection from 
$I$ into $O$. In the case of a bi-flow, $I$ and $O$ are therefore in bijection, and since one knows already that patterns with flows realize unitary embeddings, it follows that patterns with bi-flow implement unitaries.

Interestingly, one can define directly the adjoint of a pattern in the subcategory of patterns with bi-flows. Specifically, given $(f,>)$ a flow for
$(G,I,O)$, and angles $\ens{\al_i;i\in I^c}$ for preparations,
and $\ens{\ba_j;j\in O^c}$ for measurements, we write 
$\mfr P_{f,G,\vec\al,\vec\ba}$ for the pattern
obtained as in the extension to general preparations of Theorem \ref{t-flow}. 
Suppose a reverse flow $(g,>)$ is given on $(G,O,I)$, one can define:
\AR{
\mfr P_{f,G,\vec\al,\vec\ba}\dag&:=&\mfr P_{g,G,\vec\ba,\vec\al}
} 
There are two things to note here: first, for this definition to make sense, one needs to have general preparations as we described above; second, this adjunction operation depends on the choice of a reverse flow $(g,>)$. It is easy to see that $\mfr P_{f,G,\mathvec\al,\mathvec\ba}\dag$ and $\mfr P_{f,G,\mathvec\ba,\mathvec\al}$ realize adjoint unitaries. 

An example is the pattern $\mfr H:=\cx2{s_1}\m01\et12\phZ 20$ with $I=\ens1$ and $O=\ens2$. It has a unique bi-flow, and is self-adjoint in the sense that $\mfr H\dag=\mfr H$, therefore it must realize a self-adjoint operator, and indeed it realizes the Hadamard transformation.

\section{Conclusion} Whereas the one-way model had been mostly thought of in relation with the traditional circuit model, we have proposed here a flow condition, which is clearly divorced from the circuit model, and guarantees the existence of a set of Pauli corrections obtaining a (strongly and uniformly) deterministic behavior. In essence, while dealing with patterns with flow, one can wholly forget about corrections, and think of measurements as being simply projections. This in turn may help in revealing the new perspective on quantum computing which is implicit in measurement based models. Following this work, a polynomiual time algorithm for finding flow was proposed in \cite{Beaudrap06} which then extended to an algorithmic method for circuit design for unitaries thoroughly based on the one-way model \cite{BDK06}. Furthermore one can see that given an open graph state as a resource for computation, flow condition characterizes the set of all unitaries implementable on that given state. 

If one is ready to lose uniform determinism, this condition can be somewhat extended when dealing with Pauli measurements. It may be however that strong and uniform determinism is an interesting property, when it comes to fault-tolerant computing in the one-way model. 

Another point worth making is that the notion of flow gives a better understanding of why $X^\al$, $Z$ corrections and $N^\al$ preparations are needed. From the point of view of our determinism theorem (Theorem \ref{t-flow}), they represent a natural and universal way to control the non deterministic evolutions induced by 1-qubit $X-Y$ measurements on a graph state. 

Finally, although the obtained class of patterns with flow is universal, it remains to be seen whether this condition is also necessary for determinism. One also need to extend the flow condition to deal with $X-Z$ and $Y-Z$ plane measurements, which are the topics of our future work. 

\section*{Acknowledgments}

The authors wish to thank Niel de Beaudrap, Anne Broadbent, Ignacio Cirac, Paul Dumais, Damian Markham, Keiji Matsumoto for useful comments and discussions. EK was partially supported by the ARDA, MITACS, ORDCF, and CFI projects during her stay at Institute for Quantum Computing at University of Waterloo.

\end{document}